\documentclass[%
 aip,
 jmp,%
 amsmath,amssymb,
reprint,%
]{revtex4-1}

\usepackage{graphicx}
\usepackage{dcolumn}
\usepackage{bm}

\begin{document}


\title[Some coordinate transformations relevant to refractive indices]{Some coordinate transformations relevant to refractive indices}

\author{Zi-Hua Weng}
 \email{xmuwzh@xmu.edu.cn.}
\affiliation{
School of Aerospace Engineering, Xiamen University, Xiamen 361005, China\\
College of Physical Science and Technology, Xiamen University, Xiamen 361005, China
}%


\date{\today}

\begin{abstract}
  The paper focuses on applying the algebra of octonions to study some coordinate transformations in the octonion spaces, exploring the contribution of partial field potential on the speed of light. J. C. Maxwell was the first to introduce the quaternions to describe the physical properties of electromagnetic fields. Nowadays the octonions can be applied to study simultaneously the physical quantities of electromagnetic and gravitational fields, including the transformation between two coordinate systems. In the octonion space, the radius vector can be combined with the integrating function of field potential to become one composite radius vector. The latter is considered as the radius vector in an octonion composite space, which belongs to the function spaces. In the octonion composite space, when there is the relative motion between two coordinate systems, it is capable of deducing the Galilean-like transformation and Lorentz-like transformation. From the two transformations, one can achieve the influence of relative speed on the speed of light (or Sagnac effect), but also the impact of partial electromagnetic potential on the speed of light. The study states that the partial electromagnetic potential has a direct influence on the speed of light in the optical waveguides, revealing several influencing factors of refractive indices in the optical waveguides.
\end{abstract}

\maketitle


\section{Introduction}

How many types of extrinsic factors are there to vary the speed of light? Can the electromagnetic potential and others transform the speed of light, besides the electromagnetic strength and optical medium (or velocity, temperature, pressure, and density)? Since a long time ago, these puzzles have been intriguing and bewildering many scholars. For many years, the scholars strive to resolve these conundrums, revealing the reasons and influence factors to change the speed of light in the optical medium, deepening the understanding of the refractive indices. Until recently, the emergence of electromagnetic and gravitational theories, described with the algebra of octonions, is able to reply to a part of these problems. As one of theoretical applications, this field theory can be applied to explore the coordinate transformation in the octonion composite space (in Section 6). And the relevant inferences are capable of explaining why the electromagnetic potential can shift the speed of light, indicating some influence factors to change the speed of light in the optical medium.

The scholars develop some speed-of-light experiments with two light paths, validating the influence of the relative motion on the speed of light in the aether wind. These experiments are called as the two-way experiments, including the Michelson-Morley type experiment \cite{MM3,MM4,MM5}, Kennedy-Thorndike type experiment \cite{KT9,KT10,KT14}, and Ives-Stilwell type experiment \cite{IS18,IS25,OW24}. Meanwhile, some scholars explore the speed-of-light experiment with single light path, amending the deficiency of two-way experiments. This type of experiment is called as the one-way experiment \cite{OW31}. A few scholars found that several influence factors are capable of varying the speed of light, refractive index, and polarization direction, especially the electromagnetic strength \cite{EO1}, electro-optic effect \cite{EO3}, magneto-optic effect \cite{MO1,MO2,MO3}, and electromagnetically induced transparency \cite{EIT1,EIT2,EIT3}. The scholars have been searching for a few new influence factors to vary the speed of light. Some scholars find that the relative speed and meta-materials will exert an impact on the speed of light, including the Sagnac effect \cite{Sagnac1,Sagnac2,Sagnac3} and negative refractive index material \cite{NIM1,NIM3}. Meanwhile several scholars propose the varying speed of light theory \cite{varying33,varying36,varying34}.

Making a comparison and analysis of preceding studies, a few primal problems are found to be connected with the speed of light and refractive index in the classical field theory.

1) Galilean transformation. In the flat four-spacetime, the Galilean transformation claims that the speed of light is constant in the vacuum, and independent of other physical parameters. Meanwhile, the speed of light in the optical waveguide is deemed to be an intrinsic property. As a result, the Galilean transformation is unable to explain why the refractive indices are different for various optical materials. By all appearances, its scope of application is quite restricted.

2) Lorentz transformation. In the flat four-spacetime, the Lorentz transformation professes that the speed of light is an invariant constant in the vacuum. And the speed of light must be totally foreign to the relative speed and electromagnetic potential and others. However, this point of view is incapable of explaining why the refractive indices can vary in different optical waveguides. Either it is unable to answer effectively the question why the speed of light may transform within the interface layer between two different optical waveguides.

3) Measurement accuracy. During the measurement of the speed of light in the vacuum, there are not enough protection devices, to shield and eliminate the disturbance of electromagnetic fields (field strength or field potential) coming from the external circumstances. In the determination of the speed of light in the optical waveguides, except for the vacuum, the existing studies may take account of the external interference of the field strength, but neglect the intervention of the field potential.

Presenting a striking contrast to the above is that it is able to account for several puzzles, which are derived from the speed of light and refractive indices of the classical mechanics, attempting to improve the Galilean transformation and Lorentz transformation to a certain extent, in the flat spaces described with the octonions.

J. C. Maxwell was the first to introduce the quaternions to explore the physical properties of electromagnetic fields. This method edifies the subsequent scholars to apply the quaternions, octonions, and even sedenions to investigate the electromagnetic field, gravitational field, General  Theory of Relativity (GR for short), and quantum mechanics and so forth. When a part of coordinates are complex numbers, the quaternion and octonion are called as the complex-quaternion and complex-octonion respectively \cite{weng1}. Some scholars make use of the quaternions and octonions to research the electromagnetic fields \cite{chanyal1,chanyal2}, electromagnetic equations \cite{OF1,mironov1}, magnetic monopole \cite{OF2,OF3}, coordinate transformation \cite{OF4,tanisli}, optics \cite{OF5}, gravitational theory \cite{OF6,chanyal3}, GR, and quantum mechanics \cite{chanyal4,imaeda} and so on.

In the octonion spaces, the vector radius can be combined with the integrating function of field potential to become one composite vector radius of the octonion composite space (a function space). In the octonion curved composite space, it is able to deduce the relations between the physical quantity and spatial parameter. A portion of these formulae are in good accordant with the GR (see Ref.[27]). The octonion curved composite space and the GR both succeed to the Cartesian academic thought of `the space is the extension of substance (field)'. It should be noted what the paper studies is the coordinate transformation of the speed of light in the octonion flat composite space, which is an extreme circumstance of the octonion curved composite space. In the octonion flat composite space, it is capable of inferring the coordinate transformation between two coordinate systems, achieving the relations between the speed of light and the component of field potential.

1) Galilean-like transformation. In the octonion flat composite space, the Galilean-like transformation states that the speed of light is variable in the vacuum. The speed of light deals with the electromagnetic scalar potential, especially the partial electromagnetic scalar potential (in Section 7). And the speed of light is one variable physical quantity, rather than an intrinsic property any more. Consequently, it is capable of explaining why the refractive indices are different for various optical materials.

2) Lorentz-like transformation. In the octonion flat composite space, the Lorentz-like transformation reveals that the speed of light is variable in the vacuum. The speed of light associates with the partial electromagnetic scalar potential and the relative speed as well. And it is able to answer the question why the refractive indices are diverse for multifarious optical materials. As a result, it is necessary to survey the influence of partial electromagnetic scalar potential on the speed of light, in subsequent experiments.

3) Electromagnetic shielding. In the determination of the speed of light in the optical waveguides, it may encounter not only the disturbance of electromagnetic strength, but also the interference of the field potential. Therefore, during the measurement of the speed of light in the laboratories, it is essential for us to apply enough protection apparatus, to shield and eliminate the outside disturbance of electromagnetic potential, except for the intervention of the electromagnetic strength coming from the external circumstances.

The paper explores the influence of the electromagnetic potential on the speed of light in the octonion spaces. The impact of the electromagnetic potential on the speed of light is distinct from that of the existing electromagnetic strength on the speed of light. The latter comprises the electro-optic effect, magneto-optic effect, and electromagnetically induced transparency and so forth. In the experiments with respect to the influence of electromagnetic potential on the speed of light, we may refer to the Aharonov-Bohm experiment. That is, one can confine the electromagnetic strength in a certain region (such as, within the solenoid), measuring the influence of the electromagnetic potential on the speed of light in the special region, in which there is merely the electromagnetic potential but not the electromagnetic strength. It will be beneficial to further understand the physical properties of refractive indices.

\begin{table}[h]
\caption{Some subspaces and physical quantities in the electromagnetic and gravitational fields described with the octonions.}
\begin{ruledtabular}
\begin{tabular}{lll}
subspace                        &  physical quantity                                                              &  basis vector                     \\
\hline
$\mathbb{H}_g$                  &  $\mathbb{R}_g$ , $\mathbb{V}_g$ , $\mathbb{Y}_g$ , $\mathbb{A}_g$              &  $\mathbb{H}_g (\textbf{i}_j)$    \\
$\mathbb{H}_{em}$               &  $\mathbb{R}_{em}$ , $\mathbb{V}_{em}$ , $\mathbb{Y}_{em}$ , $\mathbb{A}_{em}$  &  $\mathbb{H}_{em} (\textbf{I}_j)$ \\
$\mathbb{H}_e$                  &  $\mathbb{R}_e$ , $\mathbb{V}_e$ , $\mathbb{Y}_e$ , $\mathbb{A}_e$              &  $\mathbb{H}_e (\textbf{i}_j)$    \\
\end{tabular}
\end{ruledtabular}
\end{table}

\section{Octonion space}

The octonion space $\mathbb{O}$ is able to explain the physical properties of gravitational and electromagnetic fields simultaneously. Further the octonion space can be separated into a few subspaces independent of each other, including $\mathbb{H}_g$ and $\mathbb{H}_{em}$ . The subspace $\mathbb{H}_g$ is one quaternion space, which can be applied to depict the physical properties of gravitational fields. Meanwhile, the second subspace $\mathbb{H}_{em}$ may be utilized to describe the physical properties of electromagnetic fields (Table 1).

The partial octonion $\mathbb{Y}_{em}$ , in the second subspace $\mathbb{H}_{em}$, can be rewritten as, $\mathbb{Y}_{em} = \mathbb{Y}_e \circ \textbf{I}_0$, with $\textbf{I}_0^2 = -1$ . The symbol $\circ$ denotes the octonion multiplication. The physical quantity $\mathbb{Y}_e$ can be considered as one quaternion in the second quaternion space $\mathbb{H}_e$ also. Obviously the second quaternion space $\mathbb{H}_e$ is independent of the quaternion space $\mathbb{H}_g$ . And the quaternion $\mathbb{Y}_e$ , in the second quaternion space $\mathbb{H}_e$ , is independent of the quaternion $\mathbb{Y}_g$ in the quaternion space $\mathbb{H}_g$ .

In the subspace $\mathbb{H}_g (\textbf{i}_j)$ for the gravitational fields, the coordinates are $i r_0$ and $r_k$, the basis vector is $\textbf{i}_j$ , the radius vector is, $\mathbb{R}_g = i \textbf{i}_0 r_0 + \Sigma \textbf{i}_k r_k$ , and the velocity is, $\mathbb{V}_g = i \textbf{i}_0 v_0 + \Sigma \textbf{i}_k v_k$ . Similarly, in the second subspace $\mathbb{H}_{em} (\textbf{I}_j)$ for the electromagnetic fields, the coordinates are $i R_0$ and $R_k$ , the basis vector is $\textbf{I}_j$ , the radius vector is, $\mathbb{R}_{em} = i \textbf{I}_0 R_0 + \Sigma \textbf{I}_k R_k$ , and the velocity is, $\mathbb{V}_{em} = i \textbf{I}_0 V_0 + \Sigma \textbf{I}_k V_k$ . Herein $v_0$ is the speed of graviton for the gravitational fields, while $V_0$ is the speed of photon for the electromagnetic fields. $r_0 = v_0 t$, with $t$ being the time. $r_j$ , $R_j$ , $v_j$ , and $V_j$ are all real. $\textbf{i}_0 = 1$. $\textbf{i}_k^2 = - 1$. $\textbf{I}_k^2 = - 1$. $\textbf{I}_j = \textbf{i}_j \circ \textbf{I}_0$ . $i$ is the imaginary unit. $j = 0, 1, 2, 3$. $k = 1, 2, 3$.

In the octonion space $\mathbb{O}$ , the basis vectors are $\textbf{i}_j$ and $\textbf{I}_j$ , the octonion radius vector is, $\mathbb{R} = \mathbb{R}_g + k_{eg} \mathbb{R}_{em}$ . The octonion velocity is, $\mathbb{V} = \mathbb{V}_g + k_{eg} \mathbb{V}_{em}$ . As one combination of the electric charge and mass, the charged particle possesses the octonion velocity, $\textbf{V}_{(g,em)} = \textbf{V}_g + k_{eg} \textbf{V}_{em}$. For the charged particles (Table 2), the magnitude of $\textbf{V}_{em}$ is approximately equal to that of $\textbf{V}_g$ in general, according to the force equilibrium equation \cite{weng2}. Meanwhile, as a combination of the graviton and photon, the `ordinary photon' owns the octonion velocity, $i \textbf{V}_{(0,0)} = i \textbf{i}_0 v_0 + i k_{eg} \textbf{I}_0 V_0$. And the magnitude of $V_0$ is approximately equal to that of $v_0$ in general, according to the definition of energy (see Ref.[41]). Herein $k_{eg}$ is a coefficient. $\textbf{V}_g = \Sigma \textbf{i}_k v_k$ . $\textbf{V}_{em} = \Sigma \textbf{I}_k V_k$ .

\begin{table}[h]
\caption{Several particles and velocities in the electromagnetic and gravitational fields described with the octonions.}
\begin{ruledtabular}
\begin{tabular}{@{}llc@{}}
particle              &  velocity                                                                     &    space                 \\
\hline
charged~particle      &  $\textbf{V}_{(g,em)} = \textbf{V}_g + k_{eg} \textbf{V}_{em}$                &    $\mathbb{O}$          \\
ordinary photon       &  $i \textbf{V}_{(0,0)} = i \textbf{i}_0 v_0 + i k_{eg} \textbf{I}_0 V_0$      &    $\mathbb{O}$          \\
\end{tabular}
\end{ruledtabular}
\end{table}

\section{Coordinate transformation}

In the octonion space $\mathbb{O}$ , any octonion physical quantity may own eight independent components, including one scalar quantity and seven vector quantities. When the octonion physical quantity transforms from one coordinate system $\zeta$ into another coordinate system $\eta$ , the vector quantities of this physical quantity will vary accordingly, while the scalar quantity of this physical quantity keeps unchanged.

By all appearances, the octonion radius vector $\mathbb{R}$ can transform from the coordinate system $\alpha$ into another coordinate system $\beta$ . In the coordinate system $\alpha$ , the octonion radius vector is,
\begin{eqnarray}
\mathbb{R} = i \textbf{i}_0 r_0 + \Sigma \textbf{i}_k r_k + k_{eg} ( i \textbf{I}_0 R_0 + \Sigma \textbf{I}_k R_k )   ~,
\end{eqnarray}
while this octonion radius vector in the coordinate system $\beta$ is written as,
\begin{eqnarray}
\mathbb{R}' = i \textbf{i}_0' r_0' + \Sigma \textbf{i}_k' r_k' + k_{eg} ( i \textbf{I}_0' R_0' + \Sigma \textbf{I}_k' R_k' )   ~.
\end{eqnarray}

Similarly, the octonion velocity $\mathbb{V}$ can also transform from the coordinate system $\alpha$ into the coordinate system $\beta$. In the coordinate system $\alpha$ , the octonion velocity, $\mathbb{V} = \partial \mathbb{R} / \partial t$ , is written as,
\begin{eqnarray}
\mathbb{V} = i \textbf{i}_0 v_0 + \Sigma \textbf{i}_k v_k + k_{eg} ( i \textbf{I}_0 V_0 + \Sigma \textbf{I}_k V_k )   ~,
\end{eqnarray}
while this octonion velocity in the coordinate system $\beta$ is,
\begin{eqnarray}
\mathbb{V}' = i \textbf{i}_0' v_0' + \Sigma \textbf{i}_k' v_k' + k_{eg} ( i \textbf{I}_0' V_0' + \Sigma \textbf{I}_k' V_k' )   ~.
\end{eqnarray}

In the vacuum without any external field, it is able to deduce some invariants from the above coordinate transformations with respect to the octonion radius vector and velocity.

\subsection{Invariants}

In the octonion space $\mathbb{O}$ , it is capable of inferring a few invariants, from the coordinate transformations of the octonion radius vector and velocity, including the scalar part of velocity, scalar part of radius vector, and norm of radius vector and so forth.

Firstly, when the octonion radius vector $\mathbb{R}$ transforms from the coordinate system $\alpha$ into the coordinate system $\beta$, the scalar part of the octonion radius vector $\mathbb{R}$ is an invariant under the coordinate transformations. As a result, from Eqs.(1) and (2), we achieve one inference,
\begin{eqnarray}
r_0' = r_0  ~ ,
\end{eqnarray}
where $\textbf{i}_0' = 1$.

Secondly, the norm of octonion radius vector $\mathbb{R}$ is a scalar, and also an invariant under the coordinate transformations. From Eqs.(1) and (2), there is an invariant,
\begin{eqnarray}
( \mathbb{R}' )^\ast \circ \mathbb{R}' = \mathbb{R}^\ast \circ \mathbb{R}  ~ ,
\end{eqnarray}
where $\ast$ denotes the octonion conjugate. $k_{eg}$ is supposed to be an invariant under the coordinate transformations.

Thirdly, when the octonion velocity $\mathbb{V}$ transforms from the coordinate system $\alpha$ into the coordinate system $\beta$ , the scalar part of the octonion velocity $\mathbb{V}$ is an invariant under the coordinate transformations. Consequently, from Eqs.(3) and (4), it is able to conclude one invariant,
\begin{eqnarray}
v_0' = v_0  ~ .
\end{eqnarray}

In the vacuum without any external field, a few combinations of the preceding invariants can be considered as the basic postulates of several coordinate transformations, including the Galilean transformation and Lorentz transformation (Table 3).

\subsection{Galilean transformation}

For the sake of simplicity, one may choose the basis vectors, $\textbf{i}_j'$ and $\textbf{I}_j'$, to be parallel to $\textbf{i}_j$ and $\textbf{I}_j$ respectively, in the octonion space $\mathbb{O}$. Therefore, from Eqs.(5) and (7), it is capable of deducing the two basic postulates of the Galilean transformation as follows,
\begin{eqnarray}
r_0' = r_0  ~ , ~~ v_0' = v_0  ~ ,
\end{eqnarray}
or
\begin{eqnarray}
t' = t  ~ , ~~ v_0' = v_0  ~ .
\nonumber
\end{eqnarray}

From these two basic postulates, it is able to deduce the Galilean transformation in the octonion space (in Section 4). If the octonion space $\mathbb{O}$ is degenerated into the quaternion space $\mathbb{H}_g$ , the Galilean transformation in the octonion space will be reduced into that in the classical field theory.

\subsection{Lorentz transformation}

We can select the basis vectors, $\textbf{i}_j'$ and $\textbf{I}_j'$ , to be parallel to $\textbf{i}_j$ and $\textbf{I}_j$ respectively, In the octonion space $\mathbb{O}$ . Consequently, from Eqs.(6) and (7), it is able to infer the two basic postulates of the Lorentz transformation,
\begin{eqnarray}
( \mathbb{R}' )^\ast \circ \mathbb{R}' = \mathbb{R}^\ast \circ \mathbb{R}  ~ , ~~ v_0' = v_0  ~ .
\end{eqnarray}

From these two basic postulates, one may deduce the Lorentz transformation in the octonion space (in Section 5).

In case the scalar part of octonion radius vector plays a major role, the two basic postulates, Eq.(9), will be simplified into the two basic postulates, Eq.(8), in the octonion space. If the octonion space $\mathbb{O}$ is degenerated into the quaternion space $\mathbb{H}_g$ , the two basic postulates, Eq.(9), in the octonion space is able to be reduced into the two basic postulates in the Special Theory of Relativity (Table 4).

\begin{table}[h]
\caption{Some invariants and physical quantities in the electromagnetic and gravitational fields described with the octonions.}
\begin{ruledtabular}
\begin{tabular}{@{}lll@{}}
space                        &  invariant                                                                        &    physical~quantity             \\
\hline
$\mathbb{O}$                 &  $r_0' = r_0 ~, $                                                                 &    radius~vector                 \\
                             &  $( \mathbb{R}' )^\ast \circ \mathbb{R}' = \mathbb{R}^\ast \circ \mathbb{R} ~,$   &    radius~vector                 \\
                             &  $v_0' = v_0 ~, $                                                                 &    velocity                      \\
$\mathbb{O}_U$               &  $u_0' = u_0 ~, $                                                                 &    composite~radius~vector       \\
                             &  $( \mathbb{U}' )^\ast \circ \mathbb{U}' = \mathbb{U}^\ast \circ \mathbb{U} ~,$   &    composite~radius~vector       \\
                             &  $g_0' = g_0 ~, $                                                                 &    composite~velocity            \\
\end{tabular}
\end{ruledtabular}
\end{table}

\section{Galilean transformation}

In the octonion space $\mathbb{O}$ , we choose the basis vectors of two coordinate systems, $\alpha (r_j , R_j)$ and $\beta (r_j' , R_j')$ , to be parallel to each other respectively. That is, the basis vectors, $\textbf{i}_j'$ and $\textbf{I}_j'$ , of coordinate system $\beta (r_j' , R_j')$ are respectively parallel to the basis vectors, $\textbf{i}_j$ and $\textbf{I}_j$ , of coordinate system $\alpha (r_j , R_j)$. When there is one relative velocity between two coordinate systems, $\alpha (r_j , R_j)$ and $\beta (r_j' , R_j')$ , it is able to derive the Galilean transformation for the octonion radius vector, $\mathbb{R} = \mathbb{R}_g +k_{eg} \mathbb{R}_{em}$, from the invariants, Eq.(8). In essence, it belongs to the translation transformation. In some conditions, there are some simple cases of the Galilean transformation.

The octonion field potential, $\mathbb{A} = i \lozenge^\times \circ \mathbb{X}$ , can be separated into, $\mathbb{A} = \mathbb{A}_g + k_{eg} \mathbb{A}_{em}$. Herein $\mathbb{A}_g$ and $\mathbb{A}_{em}$ are respectively the field potential in the subspace $\mathbb{H}_g$ and second subspace $\mathbb{H}_{em}$ . $\times$ denotes the complex conjugate. The quaternion operator is, $\lozenge = i \partial / \partial r_0 + \Sigma \textbf{i}_k \partial / \partial r_k$ .

Taking the octonion field potential $\mathbb{A}$ as an example, it is able to discuss some transformations of one octonion physical quantity under the Galilean transformation. It should be noted that the octonion field potential $\mathbb{A}$ is irrelevant to any coordinate $R_j$ of the radius vector $\mathbb{R}_{em}$, according to the definition of quaternion operator $\lozenge$. Only the relative velocity in the quaternion space $\mathbb{H}_g$ can make contribution to the Galilean transformation of the octonion field potential.

\subsection{Galilean transformation 1}

Taking the transformation of coordinate $r_1$ as an example, we study the Galilean transformation between two coordinates, $r_k$ and $r_k'$ . It is able to derive the Galilean transformation from the invariants, Eq.(8), when there is merely a relative velocity in the basis vector $\textbf{i}_1$ , between the two coordinate systems, $\alpha (r_j , R_j)$ and $\beta (r_j' , R_j')$.

In the octonion space $\mathbb{O}$ , the Galilean transformation for the coordinates, $r_j$ and $R_j$, of the octonion radius vector, $\mathbb{R}$ , can be written as follows,
\begin{eqnarray}
r_1' && = r_1 - v_{1(r)} t ~,
\\
r_m' && = r_m  ~~~~~ (m = 0, 2, 3) ~,
\\
R_j' && = R_j  ~~~~~ (j = 0, 1, 2, 3) ~,
\end{eqnarray}
where $v_{1(r)}$ is the relative difference of the velocity component $v_1$ , for the direction $\textbf{i}_1$ of the two coordinate systems, $\alpha (r_j , R_j)$ and $\beta (r_j' , R_j')$ . Obviously, only the coordinate $r_1$ is transformed.

\subsection{Galilean transformation 2}

In case there is merely a relative velocity in the basis vector $\textbf{I}_0$ , between the two coordinate systems, $\alpha (r_j , R_j)$ and $\beta (r_j' , R_j')$ , one can derive the Galilean transformation from the invariants, Eq.(8).

In the octonion space $\mathbb{O}$ , the Galilean transformation for the coordinates, $r_j$ and $R_j$, of the octonion radius vector, $\mathbb{R}$ , can be written as follows,
\begin{eqnarray}
r_j' && = r_j ~~~~~  (j = 0, 1, 2, 3) ~,
\\
R_0' && = R_0 - V_{0(r)} t ~,
\\
R_k' && = R_k  ~~~~~ (k = 1, 2, 3) ,
\end{eqnarray}
where $V_{0(r)}$ is the relative difference of the velocity component $V_0$ , for the direction $\textbf{I}_0$ of the two coordinate systems, $\alpha (r_j , R_j)$ and $\beta (r_j' , R_j')$ . Apparently, only the coordinate $R_0$ is transformed.

\subsection{Galilean transformation 3}

Taking the transformation of coordinate $R_1$ as an example, we study the Galilean transformation between two coordinates, $R_k$ and $R_k'$ . One can derive the Galilean transformation from the invariants, Eq.(8), when there is merely a relative velocity in the basis vector $\textbf{I}_1$ , between the two coordinate systems, $\alpha (r_j , R_j)$ and $\beta (r_j' , R_j')$.

In the octonion space $\mathbb{O}$ , the Galilean transformation for the coordinates, $r_j$ and $R_j$, of the octonion radius vector, $\mathbb{R}$ , can be written as follows,
\begin{eqnarray}
r_j' && = r_j  ~~~~~ (j = 0, 1, 2, 3) ~,
\\
R_1' && = R_1 - V_{1(r)} t ~,
\\
R_m' && = R_m ~~~~~  (m = 0, 2, 3) ~,
\end{eqnarray}
where $V_{1(r)}$ is the relative difference of the velocity component $V_1$ , for the direction $\textbf{I}_1$ of the two coordinate systems, $\alpha (r_j , R_j)$ and $\beta (r_j' , R_j')$ . Obviously, only the coordinate $R_1$ is transformed.

\subsection{Gravitational potential}

When there is merely a relative velocity in the basis vector $\textbf{i}_1$ , between the two coordinate systems, $\alpha (r_j , R_j)$ and $\beta (r_j' , R_j')$ , it is capable of deriving the Galilean transformation from the invariants, Eq.(8). The quaternion radius vector, $\mathbb{R}_g$ , abides by the Galilean transformation.

In case the electromagnetic potential $\mathbb{A}_{em}$ is equal to zero, the octonion field potential, $\mathbb{A}$ , will be reduced into the gravitational potential $\mathbb{A}_g$ . The latter possesses the same basis vector, $\textbf{i}_j$ , as the quaternion radius vector $\mathbb{R}_g$ . As a result, the transformation law of gravitational potential $\mathbb{A}_g$ is identical with that of quaternion radius vector $\mathbb{R}_g$ . That is, the gravitational potential $\mathbb{A}_g$ obeys the Galilean transformation.

\subsection{Electromagnetic potential}

One can derive the Galilean transformation from the invariants, Eq.(8), in case there is merely a relative velocity in the basis vector $\textbf{i}_1$ , between the two coordinate systems, $\alpha (r_j , R_j)$ and $\beta (r_j' , R_j')$ . The quaternion radius vector, $\mathbb{R}_g$, obeys the Galilean transformation. When there is merely the electromagnetic potential $\mathbb{A}_{em}$, it is able to transform equivalently the electromagnetic potential from the second subspace $\mathbb{H}_{em}$ into the second quaternion space $\mathbb{H}_e$ , by virtue of the relationship between the second quaternion space $\mathbb{H}_e$ and second subspace $\mathbb{H}_{em}$ .

If the gravitational potential $\mathbb{A}_g$ is equal to zero, the octonion field potential, $\mathbb{A}$ , will be reduced into the electromagnetic potential $\mathbb{A}_{em}$ . Because there exists a relative motion between the two coordinate systems, $\alpha (r_j , R_j)$ and $\beta (r_j' , R_j')$ , the electromagnetic potential $\mathbb{A}_{em}$ in the coordinate system $\alpha (r_j , R_j)$ can be transformed into the electromagnetic potential $\mathbb{A}_{em}'$ in the coordinate system $\beta (r_j' , R_j')$ , by means of the Galilean transformation. That is, $\mathbb{A}_{em} \rightarrow \mathbb{A}_{em}'$ . Multiplying both sides of this Galilean transformation by $\textbf{I}_0^\ast$ will produce, $\mathbb{A}_{em} \circ \textbf{I}_0^\ast \rightarrow \mathbb{A}_{em}' \circ \textbf{I}_0^\ast$ . It's worth noting that the basis vector $\textbf{I}_0'$ is parallel to $\textbf{I}_0$ , there are, $\textbf{I}_0' \circ \textbf{I}_0^\ast = 1$, and $\textbf{I}_0 \circ \textbf{I}_0^\ast = 1$. Consequently, the basis vectors, $\textbf{I}_j$ and $\textbf{I}_j'$, of the electromagnetic potential have been respectively transferred into the basis vectors, $\textbf{i}_j$ and $\textbf{i}_j'$ , in the Galilean transformation.

The above means that the two transferred electromagnetic potentials, $\{ \mathbb{A}_{em} \circ \textbf{I}_0^\ast \} (\textbf{i}_j)$ and $\{ \mathbb{A}_{em}' \circ \textbf{I}_0^\ast \} (\textbf{i}_j')$ , situate in the two second quaternion spaces, $\mathbb{H}_e (\textbf{i}_j)$ and $\mathbb{H}_e (\textbf{i}_j')$ , respectively. In other words, the electromagnetic potential, $\mathbb{A}_{em} (\textbf{I}_j)$ , with the basis vector $\textbf{I}_j$ has been transferred into the electromagnetic potential, $\{ \mathbb{A}_{em} \circ \textbf{I}_0^\ast \} (\textbf{i}_j)$, with the basis vector $\textbf{i}_j$ . As a result, the electromagnetic potential, $\{ \mathbb{A}_{em} \circ \textbf{I}_0^\ast \} (\textbf{i}_j)$, possesses the same basis vector, $\textbf{i}_j$ , as the radius vector $\mathbb{R}_e$ . The transformation law of electromagnetic potential, $\{ \mathbb{A}_{em} \circ \textbf{I}_0^\ast \} (\textbf{i}_j)$ , is identical with that of quaternion radius vector $\mathbb{R}_g$ . That is, the electromagnetic potential, $\{ \mathbb{A}_{em} \circ \textbf{I}_0^\ast \} (\textbf{i}_j)$ , abides by the Galilean transformation.

\begin{table}[h]
\caption{A few simple coordinate transformations and relevant basic postulates in the octonion spaces for the electromagnetic and gravitational fields.}
\begin{ruledtabular}
\begin{tabular}{ll}
coordinate~transformation     &  basic~postulate                                                                                            \\
\hline
Galilean~transformation       &  $r_0' = r_0  ~ , ~~ v_0' = v_0   ;$                                                                        \\
Lorentz~transformation        &  $( \mathbb{R}' )^\ast \circ \mathbb{R}' = \mathbb{R}^\ast \circ \mathbb{R}, $    $v_0' = v_0 ;$            \\
Galilean-like~transformation  &  $u_0' = u_0  ~ , ~~ g_0' = g_0   ;$                                                                        \\
Lorentz-like~transformation   &  $( \mathbb{U}' )^\ast \circ \mathbb{U}' = \mathbb{U}^\ast \circ \mathbb{U},$     $g_0' = g_0 ;$            \\
\end{tabular}
\end{ruledtabular}
\end{table}

\section{Lorentz transformation}

In the octonion space $\mathbb{O}$ , it is able to choose the basis vectors of two coordinate systems, $\alpha (r_j , R_j)$ and $\beta (r_j' , R_j')$, to be parallel to each other respectively. That is, the basis vectors, $\textbf{i}_j'$ and $\textbf{I}_j'$ , of coordinate system $\beta (r_j' , R_j')$ are respectively parallel to the basis vectors, $\textbf{i}_j$ and $\textbf{I}_j$ , of coordinate system $\alpha (r_j , R_j)$. When there exists a relative motion between two coordinate systems, $\alpha (r_j , R_j)$ and $\beta (r_j' , R_j')$, one may derive the Lorentz transformation for the octonion radius vector, $\mathbb{R}$ , from the invariants, Eq.(9). In essence, it belongs to the rotation transformation. In some conditions, there are several simple cases of the Lorentz transformation.

Taking the octonion field potential $\mathbb{A}$ as an example, it is able to research some transformations of one octonion physical quantity under the Lorentz transformation. It is a remarkable fact that the octonion field potential $\mathbb{A}$ is independent of any coordinate $R_j$ of the radius vector $\mathbb{R}_{em}$ , according to the definition of quaternion operator $\lozenge$. Only the relative motion in the quaternion space $\mathbb{H}_g$ can exert an impact on the Lorentz transformation of the octonion field potential.

\subsection{Lorentz transformation 1}

Taking the transformation of coordinate $r_1$ as an example, one can explore the Lorentz transformation between two coordinates, $r_0$ and $r_k$ . It is capable of deriving the Lorentz transformation from the invariants in Eq.(9), when there is only a relative velocity in the basis vector $\textbf{i}_1$ , between the two coordinate systems, $\alpha (r_j , R_j)$ and $\beta (r_j' , R_j')$ .

In the octonion space $\mathbb{O}$ , the Lorentz transformation for the coordinates, $r_j$ and $R_j$, of the octonion radius vector, $\mathbb{R}$ , can be written as follows,
\begin{eqnarray}
r_0' && = \{ r_0 - \lambda_{(v1)} r_1 \} / \eta ~ ,
\\
r_1'&& = \{ r_1 - v_{1(r)} t \} / \eta ~ ,
\\
r_d' && = r_d ~~~~~ (d = 2, 3) ~,
\\
R_j' && = R_j ~~~~~ (j = 0, 1, 2, 3) ~,
\end{eqnarray}
where $\lambda_{(v1)} = v_{1(r)} / v_0$ , $\eta = ( 1 - \lambda_{(v1)}^2 )^{1/2}$ . $v_{1(r)}$ is the relative difference of the velocity component $v_1$, for the direction $\textbf{i}_1$ of the two coordinate systems, $\alpha (r_j , R_j)$ and $\beta (r_j' , R_j')$ . In terms of the simplified invariant, $r_0'^2 - r_1'^2 = r_0^2 - r_1^2$ , only two coordinates, $r_0$ and $r_1$ , are transformed.

\subsection{Lorentz transformation 2}

If there is merely a relative motion in the basis vector $\textbf{I}_0$ , between the two coordinate systems, $\alpha (r_j , R_j)$ and $\beta (r_j' , R_j')$ , we can derive the Lorentz transformation from the invariants in Eq.(9).

In the octonion space $\mathbb{O}$ , the Lorentz transformation for the coordinates, $r_j$ and $R_j$, of the octonion radius vector, $\mathbb{R}$ , can be written as follows,
\begin{eqnarray}
r_0' && = \{ r_0 - \lambda_{(V0)} f R_0 \} / \eta ~  ,
\\
f R_0' && = \{ f R_0 - V_{0(r)} t \} / \eta ~  ,
\\
r_k' && = r_k  ~~~~~  (k = 1, 2, 3) ~,
\\
R_k' && = R_k  ~~~~~  (k = 1, 2, 3) ~,
\end{eqnarray}
where $f = i k_{eg}$ , and it is real. $k_{eg}^2 = \mu_g / \mu_e$ . $\mu_g < 0$, and $\mu_e > 0$. $\lambda_{(V0)} = V_{0(r)} / v_0$ , $\eta = ( 1 - \lambda_{(V0)}^2 )^{1/2}$ . $V_{0(r)}$ is the relative difference of the velocity component $V_0$ , for the direction $\textbf{I}_0$ of the two coordinate systems, $\alpha (r_j , R_j)$ and $\beta (r_j' , R_j')$ . In terms of the simplified invariant, $r_0'^2 - (f R_0')^2 = r_0^2 - (f R_0)^2$ , only two coordinates, $r_0$ and $R_0$, are transformed.

\subsection{Lorentz transformation 3}

Taking the transformation of coordinate $R_1$ as an example, one can study the Lorentz transformation between two coordinates, $r_0$ and $R_k$ . It is able to derive the Lorentz transformation from the invariants in Eq.(9), when there is merely a relative motion in the basis vector $\textbf{I}_1$ , between the two coordinate systems, $\alpha (r_j , R_j)$ and $\beta (r_j' , R_j')$ .

In the octonion space $\mathbb{O}$ , the Lorentz transformation for the coordinates, $r_j$ and $R_j$, of the octonion radius vector, $\mathbb{R}$ , can be written as follows,
\begin{eqnarray}
r_0' && = \{ r_0 + \lambda_{(V1)} f R_1 \} / \xi ~  ,
\\
f R_1' && = \{ f R_1 + V_{1(r)} t \} / \xi ~  ,
\\
r_k' && = r_k  ~~~~~ (k = 1, 2, 3) ~,
\\
R_m' && = R_m  ~~~~~ (m = 0, 2, 3) ~,
\end{eqnarray}
where $\lambda_{(V1)} = V_{1(r)} / v_0$ , $\xi = ( 1 + \lambda_{(V1)}^2 )^{1/2}$ . $V_{1(r)}$ is the relative difference of the velocity component $V_1$ , for the direction $\textbf{I}_1$ of the two coordinate systems, $\alpha (r_j , R_j)$ and $\beta (r_j' , R_j')$ . In terms of the simplified invariant, $r_0'^2 + (f R_1')^2 = r_0^2 + (f R_1)^2$ , only two coordinates, $r_0$ and $R_1$, are transformed.

\subsection{Gravitational potential}

When there is merely a relative motion in the basis vector $\textbf{i}_1$ , between the two coordinate systems, $\alpha (r_j , R_j)$ and $\beta (r_j' , R_j')$ , it is able to derive the Lorentz transformation from the invariants in Eq.(9). The quaternion radius vector, $\mathbb{R}_g$ , abides by the Lorentz transformation.

If the electromagnetic potential $\mathbb{A}_{em}$ is equal to zero, the octonion field potential, $\mathbb{A}$, will be degenerated into the gravitational potential $\mathbb{A}_g$ . Obviously the latter possesses the same basis vector, $\textbf{i}_j$ , as the quaternion radius vector $\mathbb{R}_g$ . Consequently the transformation law of gravitational potential $\mathbb{A}_g$ will be identical with that of quaternion radius vector $\mathbb{R}_g$ . That is, the gravitational potential $\mathbb{A}_g$ obeys the Lorentz transformation.

\subsection{Electromagnetic potential}

It is able to derive the Lorentz transformation from the invariants, Eq.(9), when there is merely a relative motion in the basis vector $\textbf{i}_1$ , between the two coordinate systems, $\alpha (r_j , R_j)$ and $\beta (r_j' , R_j')$ . The quaternion radius vector, $\mathbb{R}_g$, obeys the Lorentz transformation. When there is merely the electromagnetic potential $\mathbb{A}_{em}$, it is able to transform equivalently the electromagnetic potential from the second subspace $\mathbb{H}_{em}$ into the second quaternion space $\mathbb{H}_e$ , making use of the relationship between the second quaternion space $\mathbb{H}_e$ and second subspace $\mathbb{H}_{em}$ .

If the gravitational potential $\mathbb{A}_g$ is equal to zero, the octonion field potential, $\mathbb{A}$, will be degenerated into the electromagnetic potential $\mathbb{A}_{em}$ . Because there exists a relative motion between the two coordinate systems, $\alpha (r_j , R_j)$ and $\beta (r_j' , R_j')$, the electromagnetic potential $\mathbb{A}_{em}$ in the coordinate system $\alpha (r_j , R_j)$ can be transformed into the electromagnetic potential $\mathbb{A}_{em}'$ in the coordinate system $\beta (r_j' , R_j')$, by means of the Lorentz transformation. That is, $\mathbb{A}_{em} \rightarrow \mathbb{A}_{em}'$ . Multiplying both sides of this Lorentz transformation by $\textbf{I}_0^\ast$ will produce, $\mathbb{A}_{em} \circ \textbf{I}_0^\ast \rightarrow \mathbb{A}_{em}' \circ \textbf{I}_0^\ast$ . It's worth noting that the basis vector $\textbf{I}_0'$ is parallel to $\textbf{I}_0$ , there are, $\textbf{I}_0' \circ \textbf{I}_0^\ast = 1$, and $\textbf{I}_0 \circ \textbf{I}_0^\ast = 1$. Therefore, the basis vectors, $\textbf{I}_j$ and $\textbf{I}_j'$, of the electromagnetic potential have been respectively transferred into the basis vectors, $\textbf{i}_j$ and $\textbf{i}_j'$ , in the Lorentz transformation.

The above states that the two transferred electromagnetic potentials, $\{ \mathbb{A}_{em} \circ \textbf{I}_0^\ast \} (\textbf{i}_j)$ and $\{ \mathbb{A}_{em}' \circ \textbf{I}_0^\ast \} (\textbf{i}_j')$ , situate in the two second quaternion spaces, $\mathbb{H}_g (\textbf{i}_j)$ and $\mathbb{H}_g (\textbf{i}_j')$, respectively. In other words, the electromagnetic potential, $\mathbb{A}_{em} (\textbf{I}_j)$ , under the basis vector $\textbf{I}_j$ has been transferred into the electromagnetic potential, $\{ \mathbb{A}_{em} \circ \textbf{I}_0^\ast \} (\textbf{i}_j)$, under the basis vector $\textbf{i}_j$ . Therefore the electromagnetic potential, $\{ \mathbb{A}_{em} \circ \textbf{I}_0^\ast \} (\textbf{i}_j)$ , owns the same basis vector, $\textbf{i}_j$ , as the radius vector $\mathbb{R}_e$ . The transformation law of electromagnetic potential, $\{ \mathbb{A}_{em} \circ \textbf{I}_0^\ast \} (\textbf{i}_j)$ , must be identical with that of quaternion radius vector $\mathbb{R}_g$ . That is, the electromagnetic potential, $\{ \mathbb{A}_{em} \circ \textbf{I}_0^\ast \} (\textbf{i}_j)$, abides by the Lorentz transformation.

\begin{table}[h]
\caption{Some composite subspaces and physical quantities in the electromagnetic and gravitational fields described with the composite octonions.}
\begin{ruledtabular}
\begin{tabular}{@{}ll@{}}
physical quantity                                                                           &  basis vector                                \\
\hline
$\mathbb{R}_{g(U)}$ , $\mathbb{V}_{g(U)}$ , $\mathbb{Y}_{g(U)}$ , $\mathbb{A}_{g(U)}$       &  $\mathbb{H}_{g(U)} (\textbf{i}_j)$       \\
$\mathbb{R}_{em(U)}$ , $\mathbb{V}_{em(U)}$ , $\mathbb{Y}_{em(U)}$ , $\mathbb{A}_{em(U)}$   &  $\mathbb{H}_{em(U)} (\textbf{I}_j)$         \\
$\mathbb{R}_{e(U)}$ , $\mathbb{V}_{e(U)}$ , $\mathbb{Y}_{e(U)}$ , $\mathbb{A}_{e(U)}$       &  $\mathbb{H}_{e(U)} (\textbf{i}_j)$       \\
\end{tabular}
\end{ruledtabular}
\end{table}

\section{Composite space}

In the octonion space, although it is capable of describing the Galilean and Lorentz transformations, these two coordinate transformations are able to neither constitute the direct relations between the speed of light and the components of external fields, nor explore the reason why the speed of light may vary in various optical mediums. However, in the composite space described with the octonions, the paper is able to establish a few relations between the speed of light with the components of external fields \cite{weng3}, deducing several inferences accordant with the refractive indices.

The octonions can be utilized to explore simultaneously the electromagnetic and gravitational fields, deducing the field potential, field strength, field source, linear momentum, angular momentum, torque, energy, force, and power and so forth. When the octonion force is equal to zero, it is able to infer the force equilibrium equation, precession equilibrium equation, current continuity equation, and mass continuity equation and so on, attempting to research the energy gradient and physical jets and so forth. The radius vector $\mathbb{R}$ and the integrating function of field potential $\mathbb{X}$ both play an important role in these studies, and their contributions cannot be ignored.

In the octonion space, the radius vector $\mathbb{R}$ and the integrating function of field potential $\mathbb{X}$ can be combined together to become the composite radius vector, $\mathbb{U} = \mathbb{R} + k_{rx} \mathbb{X}$ . The composite radius vector must be considered as a whole, enabling the electromagnetic field energy and gravitational field energy both to be included in the definition of energy (see Ref.[41]). Further, this composite radius vector may also be regarded as one radius vector of the function space. The latter is called as the octonion composite space, $\mathbb{O}_U$ , temporarily. Herein $k_{rx}$ is a coefficient, to meet the requirement of the dimensional homogeneity. The integrating function of field potential is, $\mathbb{X} = i x_0 \textbf{i}_0 + \Sigma x_k \textbf{i}_k + k_{eg} ( i X_0 \textbf{I}_0 + \Sigma X_k \textbf{I}_k )$ . $x_j$ and $X_j$ are all real.

From the standpoint of function spaces, the composite space, $\mathbb{O}_U$ , is one type of function space. As a result, one may apply the research technique relevant to the function spaces to explore the physical properties of the composite space. The function space comprises multiple physical quantities, and its coordinate may be a function in general. In other words, we can look upon each coordinate of the composite spaces as one function, which consists of the coordinate of the radius vector and that of the integrating function of field potential. One may find other connections between the radius vector and the integrating function of field potential, in the octonion composite space.

From the perspective of the composite radius vector, the integrating function of field potential and the radius vector both are essential ingredients of the composite radius vector $\mathbb{U}$ . When the contribution of the radius vector $\mathbb{R}$ can be neglected, the composite radius vector will be reduced into the integrating function of field potential. In case the contribution of the integrating function of field potential $\mathbb{X}$ can be ignored, the composite radius vector will be degenerated into the radius vector. Neither of the integrating function of field potential and the radius vector can exist independently, in general. Both of them will be connected closely, and cannot be separated.

In the octonion composite space $\mathbb{O}_U$ , in the majority of cases, there exists, $\mathbb{R} \gg k_{rx} \mathbb{X}$ , that is, $\mathbb{R} + k_{rx} \mathbb{X} \approx \mathbb{R}$. Therefore, it is not an easy thing to detect the contribution of the integrating function of field potential $\mathbb{X}$ on the composite radius vector, in general. However, as some physical quantities associated with the partial derivative of the integrating function of field potential $\mathbb{X}$ , the field potential, field strength, energy, and refractive index and others have been exerting an important impact on the field theories.

The octonion composite space $\mathbb{O}_U$ is able to explain the physical properties of gravitational and electromagnetic fields simultaneously. Further the octonion composite space can be separated into a few composite subspaces independent of each other, including $\mathbb{H}_{g(U)}$ and $\mathbb{H}_{em(U)}$. The composite subspace $\mathbb{H}_{g(U)}$ is one quaternion composite space, which can be applied to depict the physical properties of gravitational fields. Meanwhile, the second composite subspace $\mathbb{H}_{em(U)}$ may be utilized to describe the physical properties of electromagnetic fields (Table 5).

The partial composite octonion $\mathbb{Y}_{em(U)}$ , in the composite subspace $\mathbb{H}_{em(U)}$ , can be rewritten as, $\mathbb{Y}_{em(U)} = \mathbb{Y}_{e(U)} \circ \textbf{I}_0$. The physical quantity $\mathbb{Y}_{e(U)}$ can be considered as one composite quaternion in the second quaternion composite space $\mathbb{H}_{e(U)}$ also. Obviously the second quaternion composite space $\mathbb{H}_{e(U)}$ is independent of the quaternion composite space $\mathbb{H}_{g(U)}$ . And the composite quaternion $\mathbb{Y}_{e(U)}$ , in the second quaternion composite space $\mathbb{H}_{e(U)}$ , is independent of the composite quaternion $\mathbb{Y}_{g(U)}$ in the quaternion composite space $\mathbb{H}_{g(U)}$ .

Apparently, the investigation method of the Galilean transformation and Lorentz transformation can be extended from the octonion space, $\mathbb{O}$ , into the octonion composite space $\mathbb{O}_U$ .

\begin{table}[h]
\caption{Comparison of some physical quantities between the octonion space $\mathbb{O}$ and octonion composite space $\mathbb{O}_U$.}
\begin{ruledtabular}
\begin{tabular}{lll}
physics~quantity             &  space $\mathbb{O}$                                                &  composite space $\mathbb{O}_U$                             \\
\hline
basis vector                 &  $\textbf{i}_j$                                                    &  $\textbf{i}_j$                                             \\
                             &  $\textbf{I}_j$                                                    &  $\textbf{I}_j$                                             \\
coordinate	                 &  $r_j$                                                             &  $u_j = r_j + k_{rx} x_j$                                   \\
                             &  $R_j$                                                             &  $U_j = R_j + k_{rx} X_j$                                   \\
time                         &  $t$	                                                              &  $t$                                                        \\
                             &  $r_0 = v_0 t$                                                     &  $u_0 = g_0 t$                                              \\
speed                        &  $v_j$                                                             &  $g_j = v_j + k_{rx} y_j$                                   \\
                             &  $V_j$                                                             &  $G_j = V_j + k_{rx} Y_j$                                   \\
relative speed               &  $v_{k(r)}$                                                        &  $g_{k(r)} = v_{k(r)} + k_{rx} y_{k(r)}$                    \\
                             &  $V_{j(r)}$                                                        &  $G_{j(r)} = V_{j(r)} + k_{rx} Y_{j(r)}$                    \\
\end{tabular}
\end{ruledtabular}
\end{table}

\section{Composite-coordinate transformation}

In the octonion composite space $\mathbb{O}_U$ , an octonion composite physical quantity (such as, composite radius vector, or composite velocity) may possess eight independent components, including one scalar quantity and seven vector quantities. When the octonion composite physical quantity transforms from one coordinate system $\theta$ into another coordinate system $\varphi$ , the vector quantities of this composite physical quantity vary accordingly, while the scalar quantity of this composite physical quantity remains the same.

By all appearances, the octonion composite radius vector $\mathbb{U}$ can transform from the coordinate system $\mu$ into another coordinate system $\nu$ . In the coordinate system $\mu$ , the octonion composite radius vector is written as,
\begin{eqnarray}
\mathbb{U} = i u_0 \textbf{i}_0 + \Sigma u_k \textbf{i}_k + k_{eg} ( i U_0 \textbf{I}_0 + \Sigma U_k \textbf{I}_k )  ~ ,
\end{eqnarray}
while this octonion composite radius vector in the coordinate system $\nu$ is,
\begin{eqnarray}
\mathbb{U}' = i u_0' \textbf{i}_0' + \Sigma u_k' \textbf{i}_k' + k_{eg} ( i U_0' \textbf{I}_0' + \Sigma U_k' \textbf{I}_k' )  ~ ,
\end{eqnarray}
where $u_j = r_j + k_{rx} x_j$, $U_j = R_j + k_{rx} X_j$. $u_j' = r_j' + k_{rx} x_j'$, $U_j' = R_j' + k_{rx} X_j'$.

Similarly, the octonion composite velocity $\mathbb{G}$ can also transform from the coordinate system $\mu$ into the coordinate system $\nu$ . In the coordinate system $\mu$ , the octonion composite velocity, $\mathbb{G} = \partial \mathbb{U} / \partial t$ , is written as,
\begin{eqnarray}
\mathbb{G} = i g_0 \textbf{i}_0 + \Sigma g_k \textbf{i}_k + k_{eg} ( i G_0 \textbf{I}_0 + \Sigma G_k \textbf{I}_k )  ~ ,
\end{eqnarray}
while this octonion composite velocity in the coordinate system $\nu$ is,
\begin{eqnarray}
\mathbb{G}' = i g_0' \textbf{i}_0' + \Sigma g_k' \textbf{i}_k' + k_{eg} ( i G_0' \textbf{I}_0' + \Sigma G_k' \textbf{I}_k' )  ~ ,
\end{eqnarray}
where $g_j = v_j + k_{rx} y_j$, $G_j = V_j + k_{rx} Y_j$. $g_j' = v_j' + k_{rx} y_j'$, $G_j' = V_j' + k_{rx} Y_j'$.
$u_0 = g_0 t$ , $U_0 = G_0 t$ .
$\mathbb{Y} = \partial \mathbb{X} / \partial t$, and it is merely relevant to one special part of field potential. $\mathbb{Y}$ is called as the partial potential temporarily. $\mathbb{Y} = i y_0 \textbf{i}_0 + \Sigma y_k \textbf{i}_k + k_{eg} ( i Y_0 \textbf{I}_0 + \Sigma Y_k \textbf{I}_k )$. $\mathbb{Y}' = i y_0' \textbf{i}_0' + \Sigma y_k' \textbf{i}_k' + k_{eg} ( i Y_0' \textbf{I}_0' + \Sigma Y_k' \textbf{I}_k' )$ . $y_j$ , $Y_j$ , $y_j'$, and $Y_j'$ are all real. $k_{rx}$ is an invariant under the coordinate transformation. $k_{rx} = 1 / c$, with $c$ being the speed of `ordinary photon' in the vacuum without any field potential nor field strength.

In the vacuum with the nonzero external fields, it is able to infer several invariants from preceding coordinate transformations with respect to the octonion composite radius vector and velocity (Table 6).

\subsection{Scalar quantity}

In the octonion composite space $\mathbb{O}_U$ , it is able to deduce a few invariants from the octonion composite radius vector and composite velocity, including the scalar of composite velocity, scalar of composite radius vector, and norm of composite radius vector and so forth.

When the octonion composite radius vector $\mathbb{U}$ transforms from the coordinate system $\mu$ into the coordinate system $\nu$ , the scalar part of the octonion composite radius vector $\mathbb{U}$ is an invariant under the coordinate transformations. As a result, from Eqs.(31) and (32), we infer an invariant,
\begin{eqnarray}
u_0' = u_0  ~ .
\end{eqnarray}

Similarly, the norm of octonion composite radius vector $\mathbb{U}$ is a scalar, and also an invariant under the coordinate transformations. From Eqs.(31) and (32), there is one inference,
\begin{eqnarray}
( \mathbb{U}' )^\ast \circ \mathbb{U}' = \mathbb{U}^\ast \circ \mathbb{U}   ~ .
\end{eqnarray}

Furthermore, when the octonion composite velocity $\mathbb{G}$ transforms from the coordinate system $\mu$ into the coordinate system $\nu$ , the scalar part of the octonion composite velocity $\mathbb{G}$ is an invariant under the coordinate transformations. Subsequently, from Eqs.(33) and (34), it is able to achieve one invariant,
\begin{eqnarray}
g_0' = g_0  ~ .
\end{eqnarray}

In the vacuum with the external fields, some combinations of the preceding invariants can be regarded as the basic postulates of several coordinate transformations, including the Galilean-like transformation and Lorentz-like transformation.

\subsection{Galilean-like transformation}

For the sake of simplicity, one can still choose the basis vectors, $\textbf{i}_j'$ and $\textbf{I}_j'$, to be parallel to $\textbf{i}_j$ and $\textbf{I}_j$ respectively, in the octonion composite space $\mathbb{O}_U$ . Consequently, from Eqs.(35) and (37), it is capable of inferring the following basic postulates of the Galilean-like transformation,
\begin{eqnarray}
u_0' = u_0  ~ ,  ~~ g_0' = g_0  ~ .
\end{eqnarray}

From the two basic postulates, it is able to deduce the Galilean-like transformation in the octonion composite space $\mathbb{O}_U$ (in Section 8).

\subsection{Lorentz-like transformation}

We can still select the basis vectors, $\textbf{i}_j'$ and $\textbf{I}_j'$ , to be parallel to $\textbf{i}_j$ and $\textbf{I}_j$ respectively, in the octonion composite space $\mathbb{O}_U$ . As a result, from Eqs.(36) and (37), it is able to deduce the two basic postulates of the Lorentz-like transformation as follows,
\begin{eqnarray}
( \mathbb{U}' )^\ast \circ \mathbb{U}' = \mathbb{U}^\ast \circ \mathbb{U}   ~ ,  ~~ g_0' = g_0  ~ .
\end{eqnarray}

From the two basic postulates, we can derive the Lorentz-like transformation in the octonion composite space $\mathbb{O}_U$ (in Section 9).

\section{Galilean-like transformation}

In the octonion composite space $\mathbb{O}_U$ , we choose the basis vectors of two coordinate systems, $\mu (u_j , U_j)$ and $\nu (u_j' , U_j')$, to be parallel to each other respectively. That is, the basis vectors, $\textbf{i}_j'$ and $\textbf{I}_j'$ , of coordinate system $\nu (u_j' , U_j')$ are respectively parallel to the basis vectors, $\textbf{i}_j$ and $\textbf{I}_j$ , of coordinate system $\mu (u_j , U_j)$. When there is one relative velocity between two coordinate systems, $\mu (u_j , U_j)$ and $\nu (u_j' , U_j')$ , it is able to derive the Galilean-like transformation for the octonion composite radius vector, $\mathbb{U}$ , from the invariants, Eq.(38). In essence, it belongs to the translation transformation also. In some conditions, there are some simple cases of the Galilean-like transformation (Table 7).

The octonion composite radius vector can be separated into, $\mathbb{U} = \mathbb{U}_g + k_{eg} \mathbb{U}_{em}$. Meanwhile the octonion composite field potential, $\mathbb{A}_U = i \lozenge^\times \circ \mathbb{U} / k_{rx}$ , can be separated into, $\mathbb{A}_U = \mathbb{A}_{g(U)} + k_{eg} \mathbb{A}_{em(U)}$ . Herein $\mathbb{U}_g$ and $\mathbb{U}_{em}$ are respectively the composite radius vector in the composite subspace $\mathbb{H}_{g(U)}$ and second composite subspace $\mathbb{H}_{em(U)}$. And $\mathbb{A}_{g(U)}$ and $\mathbb{A}_{em(U)}$ are respectively the composite field potential in the composite subspace $\mathbb{H}_{g(U)}$ and second composite subspace $\mathbb{H}_{em(U)}$ . The operator, $\lozenge = i \partial / \partial u_0 + \Sigma \textbf{i}_k \partial / \partial u_k$ .

Taking the octonion composite field potential $\mathbb{A}_U$ as an example, it is able to explore some transformations of one octonion physical quantity under the Galilean-like transformation. It is noteworthy that the octonion composite field potential $\mathbb{A}_U$ is irrelevant to any coordinate $U_j$ of the composite radius vector $\mathbb{U}_{em}$ , according to the definition of quaternion operator $\lozenge$. Only the relative velocity in the quaternion composite space $\mathbb{H}_{g(U)}$ can make contribution to the Galilean-like transformation of the octonion composite field potential.

\subsection{Galilean-like transformation 1}

Taking the transformation of coordinate $u_1$ as an example, one can study the Galilean-like transformation between two coordinates, $u_k$ and $u_k'$ . It is able to derive the Galilean-like transformation from the invariants, Eq.(38), when there is merely a relative velocity in the basis vector $\textbf{i}_1$ , between the two coordinate systems, $\mu (u_j , U_j)$ and $\nu (u_j' , U_j')$.

In the composite octonion space $\mathbb{O}_U$ , the Galilean-like transformation for the coordinates, $u_j$ and $U_j$, of the octonion composite radius vector, $\mathbb{U}$ , can be written as follows,
\begin{eqnarray}
u_1' && = u_1 - g_{1(r)} t ~ ,
\\
u_m' && = u_m  ~~~~~ (m = 0, 2, 3) ~,
\\
U_j' && = U_j  ~~~~~ (j = 0, 1, 2, 3) ~,
\end{eqnarray}
where $g_{1(r)}$ is the relative difference of the velocity component $g_1$ , for the direction $\textbf{i}_1$ of the two coordinate systems, $\mu (u_j , U_j)$ and $\nu (u_j' , U_j')$ . Apparently, only the coordinate $u_1$ is transformed.

\subsection{Galilean-like transformation 2}

One can derive the Galilean-like transformation from the invariants, Eq.(38), in case there is merely a relative velocity in the basis vector $\textbf{I}_0$ , between the two coordinate systems, $\mu (u_j , U_j)$ and $\nu (u_j' , U_j')$ .

In the composite octonion space $\mathbb{O}_U$ , the Galilean-like transformation for the coordinates, $u_j$ and $U_j$, of the octonion composite radius vector, $\mathbb{U}$ , can be written as follows,
\begin{eqnarray}
u_j' && = u_j  ~~~~~ (j = 0, 1, 2, 3) ~,
\\
U_0' && = U_0 - G_{0(r)} t ~,
\\
U_k' && = U_k  ~~~~~ (k = 1, 2, 3) ~,
\end{eqnarray}
where $G_{0(r)}$ is the relative difference of the velocity component $G_0$ , for the direction $\textbf{I}_0$ of the two coordinate systems, $\mu (u_j , U_j)$ and $\nu (u_j' , U_j')$. Obviously, only the coordinate $U_0$ is transformed.

\subsection{Galilean-like transformation 3}

Taking the transformation of coordinate $U_1$ as an example, we may research the Galilean-like transformation between two coordinates, $U_k$ and $U_k'$ . It is able to derive the Galilean-like transformation from the invariants, Eq.(38), in case there is merely a relative velocity in the basis vector $\textbf{I}_1$ , between the two coordinate systems, $\mu (u_j , U_j)$ and $\nu (u_j' , U_j')$.

In the composite octonion space $\mathbb{O}_U$ , the Galilean-like transformation for the coordinates, $u_j$ and $U_j$, of the octonion composite radius vector, $\mathbb{U}$ , can be written as follows,
\begin{eqnarray}
u_j' && = u_j  ~~~~~ (j = 0, 1, 2, 3) ~,
\\
U_1' && = U_1 - G_{1(r)} t ~,
\\
U_m' && = U_m  ~~~~~ (m = 0, 2, 3) ~,
\end{eqnarray}
where $G_{1(r)}$ is the relative difference of the velocity component $G_1$ , for the direction $\textbf{I}_1$ of the two coordinate systems, $\mu (u_j , U_j)$ and $\nu (u_j' , U_j')$ . Apparently, only the coordinate $U_1$ is transformed.

\subsection{Composite gravitational potential}

When there is merely a relative velocity in the basis vector $\textbf{i}_1$ , between the two coordinate systems, $\mu (u_j , U_j)$ and $\nu (u_j' , U_j')$ , it is capable of deriving the Galilean-like transformation from the invariants, Eq.(38). The quaternion composite radius vector, $\mathbb{U}_g$ , abides by the Galilean-like transformation.

In case the composite electromagnetic potential $\mathbb{A}_{em(U)}$ is equal to zero, the octonion composite field potential, $\mathbb{A}_U$, will be reduced into the composite gravitational potential $\mathbb{A}_{g(U)}$ . The latter possesses the same basis vector, $\textbf{i}_j$ , as the quaternion composite radius vector $\mathbb{U}_g$ . As a result, the transformation law of composite gravitational potential $\mathbb{A}_{g(U)}$ is identical with that of quaternion composite radius vector $\mathbb{U}_g$. That is, the composite gravitational potential $\mathbb{A}_{g(U)}$ obeys the Galilean-like transformation.

\subsection{Composite electromagnetic potential}

One can derive the Galilean-like transformation from the invariants, Eq.(38), in case there is merely a relative velocity in the basis vector $\textbf{i}_1$ , between the two coordinate systems, $\mu (u_j , U_j)$ and $\nu (u_j' , U_j')$. The quaternion composite radius vector, $\mathbb{U}_g$ , obeys the Galilean-like transformation. When there is merely the composite electromagnetic potential $\mathbb{A}_{em(U)}$ , it is able to transform equivalently the composite electromagnetic potential from the second composite subspace $\mathbb{H}_{em(U)}$ into the second quaternion composite space $\mathbb{H}_{e(U)}$ , by virtue of the relationship between the second quaternion composite space $\mathbb{H}_{e(U)}$ and second composite subspace $\mathbb{H}_{em(U)}$ .

If the composite gravitational potential $\mathbb{A}_{g(U)}$ is equal to zero, the octonion composite field potential, $\mathbb{A}_U$ , will be reduced into the composite electromagnetic potential $\mathbb{A}_{em(U)}$ . Because there exists a relative motion between the two coordinate systems, $\mu (u_j , U_j)$ and $\nu (u_j' , U_j')$, the composite electromagnetic potential $\mathbb{A}_{em(U)}$ in the coordinate system $\mu (u_j , U_j)$ can be transformed into the composite electromagnetic potential $\mathbb{A}_{em(U)}'$ in the coordinate system $\nu (u_j' , U_j')$, by means of the Galilean-like transformation. That is, $\mathbb{A}_{em(U)} \rightarrow \mathbb{A}_{em(U)}'$ . Multiplying both sides of this Galilean-like transformation by $\textbf{I}_0^\ast$ will produce, $\mathbb{A}_{em(U)} \circ \textbf{I}_0^\ast \rightarrow \mathbb{A}_{em(U)}' \circ \textbf{I}_0^\ast$ . It's worth noting that the basis vector $\textbf{I}_0'$ is parallel to $\textbf{I}_0$ , there are, $\textbf{I}_0' \circ \textbf{I}_0^\ast = 1$, and $\textbf{I}_0 \circ \textbf{I}_0^\ast = 1$ . Consequently, the basis vectors, $\textbf{I}_j$ and $\textbf{I}_j'$ , of the composite electromagnetic potential have been respectively transferred into the basis vectors, $\textbf{i}_j$ and $\textbf{i}_j'$ , in the Galilean-like transformation.

The above means that the two transferred composite electromagnetic potentials, $\{ \mathbb{A}_{em(U)} \circ \textbf{I}_0^\ast \} (\textbf{i}_j)$ and $\{ \mathbb{A}_{em(U)}' \circ \textbf{I}_0^\ast \} (\textbf{i}_j')$ , situate in the two second quaternion composite spaces, $\mathbb{H}_{e(U)} (\textbf{i}_j)$ and $\mathbb{H}_{e(U)} (\textbf{i}_j')$ , respectively. In other words, the composite electromagnetic potential, $\mathbb{A}_{em(U)} (\textbf{I}_j)$ , with the basis vector $\textbf{I}_j$ has been transferred into the composite electromagnetic potential, $\{ \mathbb{A}_{em(U)} \circ \textbf{I}_0^\ast \} (\textbf{i}_j)$ , with the basis vector $\textbf{i}_j$ . As a result, the composite electromagnetic potential, $\{ \mathbb{A}_{em(U)} \circ \textbf{I}_0^\ast \} (\textbf{i}_j)$ , possesses the same basis vector, $\textbf{i}_j$ , as the composite radius vector $\mathbb{U}_e$ . The transformation law of composite electromagnetic potential, $\{ \mathbb{A}_{em(U)} \circ \textbf{I}_0^\ast \} (\textbf{i}_j)$ , is identical with that of composite radius vector $\mathbb{U}_g$ . That is, the composite electromagnetic potential, $\{ \mathbb{A}_{em(U)} \circ \textbf{I}_0^\ast \} (\textbf{i}_j)$ , abides by the Galilean-like transformation.

\section{Lorentz-like transformation}

In the octonion composite space $\mathbb{O}_U$ , it is able to select the basis vectors of two coordinate systems, $\mu (u_j , U_j)$ and $\nu (u_j' , U_j')$ , to be parallel to each other respectively. That is, the basis vectors, $\textbf{i}_j'$ and $\textbf{I}_j'$ , of coordinate system $\nu (u_j' , U_j')$ are respectively parallel to the basis vectors, $\textbf{i}_j$ and $\textbf{I}_j$ , of coordinate system $\mu (u_j , U_j)$. When there is a relative motion between two coordinate systems, $\mu (u_j , U_j)$ and $\nu (u_j' , U_j')$, one can derive the Lorentz-like transformation for the octonion composite radius vector, $\mathbb{U}$ , from the invariants in Eq.(39). In essence, it belongs to the rotation transformation also. In some conditions, there are some simple cases of the Lorentz-like transformation.

Taking the octonion composite field potential $\mathbb{A}_U$ as an example, it is able to explore some transformations of one octonion physical quantity under the Lorentz-like transformation. It is noteworthy that the octonion composite field potential $\mathbb{A}_U$ is independent of any coordinate $U_j$ of the second quaternion composite radius vector $\mathbb{U}_{em}$ , according to the definition of quaternion operator $\lozenge$. Only the relative velocity in the quaternion composite space $\mathbb{H}_{g(U)}$ may have influence on the Lorentz-like transformation of the octonion composite field potential.

\subsection{Lorentz-like transformation 1}

Taking the transformation of coordinate $u_1$ as an example, we may explore the Lorentz-like transformation between two coordinates, $u_0$ and $u_k$ . It is able to derive the Lorentz-like transformation from the invariants in Eq.(39), when there is merely a relative motion in the basis vector $\textbf{i}_1$ , between the two coordinate systems, $\mu (u_j , U_j)$ and $\nu (u_j' , U_j')$ .

In the composite octonion space $\mathbb{O}_U$ , the Lorentz-like transformation for the coordinates, $u_j$ and $U_j$ , of the octonion composite radius vector, $\mathbb{U}$ , can be written as follows,
\begin{eqnarray}
u_0' && = \{ u_0 - \lambda_{(g1)} u_1 \} / \gamma ~,
\\
u_1' && = \{ u_1 - g_{1(r)} t \} / \gamma ~,
\\
u_d' && = u_d  ~~~~~ (d = 2, 3) ~,
\\
U_j' && = U_j  ~~~~~ (j = 0, 1, 2, 3) ~,
\end{eqnarray}
where $\lambda_{(g1)} = g_{1(r)} / g_0$ , $\gamma = ( 1 - \lambda_{(g1)}^2 )^{1/2}$ . $g_{1(r)}$ is the relative difference of the velocity component $g_1$, for the direction $\textbf{i}_1$ of the two coordinate systems, $\mu (u_j , U_j)$ and $\nu (u_j' , U_j')$ . In terms of the simplified invariant, $u_0'^2 - u_1'^2 = u_0^2 - u_1^2$ , only two coordinates, $u_0$ and $u_1$ , are transformed.

\subsection{Lorentz-like transformation 2}

One can derive the Lorentz-like transformation from the invariants in Eq.(39), when there is merely a relative motion in the basis vector $\textbf{I}_0$ , between the two coordinate systems, $\mu (u_j , U_j)$ and $\nu (u_j' , U_j')$ .

In the composite octonion space $\mathbb{O}_U$ , the Lorentz-like transformation for the coordinates, $u_j$ and $U_j$ , of the octonion composite radius vector, $\mathbb{U}$ , can be written as follows,
\begin{eqnarray}
u_0' && = \{ u_0 - \lambda_{(G0)} f U_0 \} / \gamma ~ ,
\\
f U_0' && = \{ f U_0 - G_{0(r)} t \} / \gamma ~ ,
\\
u_k' && = u_k ~~~~~  (k = 1, 2, 3) ~,
\\
U_k' && = U_k ~~~~~  (k = 1, 2, 3) ~,
\end{eqnarray}
where $\lambda_{(G0)} = G_{0(r)} / g_0$ , $\gamma = ( 1 - \lambda_{(G0)}^2 )^{1/2}$ . $G_{0(r)}$ is the relative difference of the velocity component $G_0$ , for the direction $\textbf{I}_0$ of the two coordinate systems, $\mu (u_j , U_j)$ and $\nu (u_j' , U_j')$ . In terms of the simplified invariant, $u_0'^2 - (f U_0')^2 = u_0^2 - (f U_0)^2$ , only two coordinates, $u_0$ and $U_0$, are transformed.

\subsection{Lorentz-like transformation 3}

Taking the transformation of coordinate $U_1$ as an example, we can discuss the Lorentz-like transformation between two coordinates, $u_0$ and $U_k$ . One will derive the Lorentz-like transformation from the invariants in Eq.(39), when there is merely a relative motion in the basis vector $\textbf{I}_1$ , between the two coordinate systems, $\mu (u_j , U_j)$ and $\nu (u_j' , U_j')$.

In the composite octonion space $\mathbb{O}_U$ , the Lorentz-like transformation for the coordinates, $u_j$ and $U_j$ , of the octonion composite radius vector, $\mathbb{U}$ , can be written as follows,
\begin{eqnarray}
u_0' && = \{ u_0 + \lambda_{(G1)} f U_1 \} / \delta ~ ,
\\
f U_1' && = \{ f U_1 + G_{1(r)} t \} / \delta ~ ,
\\
u_k' && = u_k  ~~~~~ (k = 1, 2, 3) ~,
\\
U_m' && = U_m  ~~~~~ (m = 0, 2, 3) ~,
\end{eqnarray}
where $\lambda_{(G1)} = G_{1(r)} / g_0$ , $\delta = ( 1 + \lambda_{(G1)}^2 )^{1/2}$ . $G_{1(r)}$ is the relative difference of the velocity component $G_1$ , for the direction $\textbf{I}_1$ of the two coordinate systems, $\mu (u_j , U_j)$ and $\nu (u_j' , U_j')$ . In terms of the simplified invariant, $u_0'^2 + (f U_1')^2 = u_0^2 + (f U_1)^2$ , only two coordinates, $u_0$ and $U_1$, are transformed.

\subsection{Composite gravitational potential}

When there is merely a relative velocity in the basis vector $\textbf{i}_1$ , between the two coordinate systems, $\mu (u_j , U_j)$ and $\nu (u_j' , U_j')$ , it is capable of deriving the Lorentz-like transformation from the invariants in Eq.(39). The quaternion composite radius vector, $\mathbb{U}_g$ , abides by the Lorentz-like transformation.

If the composite electromagnetic potential $\mathbb{A}_{em(U)}$ is equal to zero, the octonion composite field potential, $\mathbb{A}_U$, will be degenerated into the composite gravitational potential $\mathbb{A}_{g(U)}$ . The latter possesses the same basis vector, $\textbf{i}_j$ , as the quaternion composite radius vector $\mathbb{U}_g$. Therefore the transformation law of composite gravitational potential $\mathbb{A}_{g(U)}$ is identical with that of quaternion composite radius vector $\mathbb{U}_g$ . That is, the composite gravitational potential $\mathbb{A}_{g(U)}$ obeys the Lorentz-like transformation.

\subsection{Composite electromagnetic potential}

It is able to derive the Lorentz-like transformation from the invariants in Eq.(39), if there is merely a relative velocity in the basis vector $\textbf{i}_1$ , between the two coordinate systems, $\mu (u_j , U_j)$ and $\nu (u_j' , U_j')$. The quaternion composite radius vector, $\mathbb{U}_g$, obeys the Lorentz-like transformation. When there is merely the composite electromagnetic potential $\mathbb{A}_{em(U)}$ , one can transform equivalently the composite electromagnetic potential from the second composite subspace $\mathbb{H}_{em(U)}$ into the second quaternion composite space $\mathbb{H}_{e(U)}$ , by means of the relationship between the second quaternion composite space $\mathbb{H}_{e(U)}$ and second composite subspace $\mathbb{H}_{em(U)}$ .

If the composite gravitational potential $\mathbb{A}_{g(U)}$ is equal to zero, the octonion composite field potential, $\mathbb{A}_U$ , will be reduced into the composite electromagnetic potential $\mathbb{A}_{em(U)}$ . Because there exists a relative motion between the two coordinate systems, $\mu (u_j , U_j)$ and $\nu (u_j' , U_j')$ , the composite electromagnetic potential $\mathbb{A}_{em(U)}$ in the coordinate system $\mu (u_j , U_j)$ can be transformed into the composite electromagnetic potential $\mathbb{A}_{em(U)}'$ in the coordinate system $\nu (u_j' , U_j')$, by means of the Lorentz-like transformation. That is, $\mathbb{A}_{em(U)} \rightarrow \mathbb{A}_{em(U)}'$ . Multiplying both sides of this Lorentz-like transformation by $\textbf{I}_0^\ast$ will produce, $\mathbb{A}_{em(U)} \circ \textbf{I}_0^\ast \rightarrow \mathbb{A}_{em(U)}' \circ \textbf{I}_0^\ast$ . It's worth noting that the basis vector $\textbf{I}_0'$ is parallel to $\textbf{I}_0$ , there are, $\textbf{I}_0' \circ \textbf{I}_0^\ast = 1$ , and $\textbf{I}_0 \circ \textbf{I}_0^\ast = 1$. Consequently, the basis vectors, $\textbf{I}_j$ and $\textbf{I}_j'$ , of the composite electromagnetic potential have been respectively transferred into the basis vectors, $\textbf{i}_j$ and $\textbf{i}_j'$ , in the Lorentz-like transformation.

The above states that the two transferred composite electromagnetic potentials, $\{ \mathbb{A}_{em(U)} \circ \textbf{I}_0^\ast \} (\textbf{i}_j)$ and $\{ \mathbb{A}_{em(U)}' \circ \textbf{I}_0^\ast \} (\textbf{i}_j')$ , situate in the two second quaternion composite spaces, $\mathbb{H}_{e(U)} (\textbf{i}_j)$ and $\mathbb{H}_{e(U)} (\textbf{i}_j')$ , respectively. In other words, the composite electromagnetic potential, $\mathbb{A}_{em(U)} (\textbf{I}_j)$ , with the basis vector $\textbf{I}_j$ has been transferred into the composite electromagnetic potential, $\{ \mathbb{A}_{em(U)} \circ \textbf{I}_0^\ast \} (\textbf{i}_j)$ , with the basis vector $\textbf{i}_j$ . As a result, the composite electromagnetic potential, $\{ \mathbb{A}_{em(U)} \circ \textbf{I}_0^\ast \} (\textbf{i}_j)$ , possesses the same basis vector, $\textbf{i}_j$ , as the composite radius vector $\mathbb{U}_e$ . The transformation law of composite electromagnetic potential, $\{ \mathbb{A}_{em(U)} \circ \textbf{I}_0^\ast \} (\textbf{i}_j)$, is identical with that of quaternion composite radius vector $\mathbb{U}_g$ . That is, the composite electromagnetic potential, $\{ \mathbb{A}_{em(U)} \circ \textbf{I}_0^\ast \} (\textbf{i}_j)$ , abides by the Lorentz-like transformation.

\begin{table}[h]
\caption{Some octonion composite physical quantities with respect to the speed of light in the electromagnetic and gravitational fields.}
\begin{ruledtabular}
\begin{tabular}{ll}
definition                                          &  constituents                                               \\
\hline
$\mathbb{U} = \mathbb{R} + k_{rx} \mathbb{X} $      &  radius~vector $\mathbb{R}$ ,                               \\
                                                    &  integrating function  $ \mathbb{X} $  ;                    \\
$\mathbb{G} = \mathbb{V} + k_{rx} \mathbb{Y} $      &  velocity $\mathbb{V}$ ,                                    \\
                                                    &  partial potential $ \mathbb{Y} $ ;                         \\
$G_0 = V_0 + k_{rx} Y_0 $                           &  speed~of~light $V_0$ ,                                     \\
                                                    &  partial scalar potential $ Y_0 $ ;                         \\
\end{tabular}
\end{ruledtabular}
\end{table}

\section{Speed of light}

In the composite octonion space $\mathbb{O}_U$, one is able to deduce a few coordinate transformations, making use of the properties of some invariants. From Eqs.(38) and (39), it is capable of inferring the Galilean-like transformation and Lorentz-like transformation and others. And they reveal that the speed of light is variable under certain circumstances.

\subsection{Galilean-like transformation}

It is supposed that there is merely a relative composite velocity in the basis vector $\textbf{i}_1$, between the two coordinate systems, $\mu$ and $\nu$ . In case this relative composite velocity is one constant $g_{1(r)}$ , the Galilean-like transformation will be written as,
\begin{eqnarray}
u_1' && = u_1 - g_{1(r)} t  ~ ,
\\
~~
t' && = t  ~ ,
\end{eqnarray}
where $g_{1(r)} = v_{1(r)} + k_{rx} y_{1(r)}$ . $v_{1(r)}$ is the relative difference of the velocity component $v_1$, while $y_{1(r)}$ is the relative difference of the component $y_1$ of the partial potential, for the direction $\textbf{i}_1$ of the two coordinate systems.

The above states that the constituents, $y_0$ and $y_1$ , of the gravitational potential will impact directly the Galilean-like transformation between the space and time, under the vacuum with nonzero external field. In case the octonion composite space $\mathbb{O}_U$ is degenerated into the octonion space $\mathbb{O}$ , the Galilean-like transformation, Eqs.(61) and (62), in the octonion composite space $\mathbb{O}_U$ will be reduced into the Galilean transformation in the octonion space $\mathbb{O}$ .

\subsection{Lorentz-like transformation}

It is supposed that there is merely a relative composite velocity in the basis vector $\textbf{i}_1$, between the two coordinate systems, $\mu$ and $\nu$ . In case this relative composite velocity is one constant $g_{1(r)}$ , we can derive the Lorentz-like transformation from Eq.(39),
\begin{eqnarray}
u_1' && = ( u_1 - g_{1(r)} t ) / \gamma ~ ,
\\
~~
u_0' && = \{ u_0 - u_1 ( g_{1(r)} / g_0 ) \} / \gamma  ~ ,
\end{eqnarray}
where $ \gamma = \{ 1 - ( g_{1(r)} / g_0 )^2 \}^{1/2} $ .

In case the scalar part of the octonion composite radius vector $\mathbb{U}$ plays a major role, the Lorentz-like transformation, Eqs.(63) and (64), will be simplified into the Galilean-like transformation, Eqs.(61) and (62), in the octonion composite space. If the octonion composite space $\mathbb{O}_U$ is degenerated into the octonion space $\mathbb{O}$, the Lorentz-like transformation, Eqs.(63) and (64), in the octonion composite space will be reduced into the Lorentz transformation in the octonion space.

\subsection{Composite speed of light}

From Eq.(39), it is capable of inferring not only the Lorentz-like transformation between the coordinates, $u_0$ and $u_k$, but also the Lorentz-like transformation between the coordinates, $u_0$ and $U_j$ , in the octonion composite space $\mathbb{O}_U$. Especially, we can achieve the coordinate transformation with respect to the speed of light, from the Lorentz-like transformation between the coordinates, $u_0$ and $U_0$ .

In the octonion composite space $\mathbb{O}_U$ , it is able to select the basis vectors, $\textbf{i}_j'$ and $\textbf{I}_j'$, to be parallel to $\textbf{i}_j$ and $\textbf{I}_j$ respectively. It is supposed that there is merely a relative composite velocity in the basis vector $\textbf{I}_0$ , between the two coordinate systems, $\mu$ and $\nu$ . If this relative composite velocity is constant, Eq.(36) will be simplified as,
\begin{eqnarray}
&& ( i \textbf{i}_0' u_0' + f \textbf{I}_0' U_0' )^\ast \circ ( i \textbf{i}_0' u_0' + f \textbf{I}_0' U_0' )
\nonumber
\\
~~
= && ( i \textbf{i}_0 u_0 + f \textbf{I}_0 U_0 )^\ast \circ ( i \textbf{i}_0 u_0 + f \textbf{I}_0 U_0 )  ~ .
\end{eqnarray}

In case this relative composite velocity, $G_{0(r)}$ , is a constant in the direction $\textbf{I}_0$, the Lorentz-like transformation for the speed of light can be derived from Eqs.(39) and (65),
\begin{eqnarray}
f U_0' && = ( f U_0 - G_{0(r)} t ) / \gamma ~ ,
\\
~~
u_0' && = \{ u_0 - f U_0 ( G_{0(r)} / g_0 ) \} / \gamma  ~ ,
\end{eqnarray}
where $\gamma = \{ 1 - ( G_{0(r)} / g_0 )^2 \}^{1/2}$ . $G_{0(r)} = V_{0(r)} + k_{rx} Y_{0(r)}$. $V_{0(r)}$ is the relative difference of the velocity component $V_0$ , while $Y_{0(r)}$ is the relative difference of the component $Y_0$ of the partial potential, for the direction $\textbf{I}_0$ of the two coordinate systems.

Further, the above may infer the coordinate transformation with respect to the composite speed of light, $G_0' = U_0' / t' $, as follows,
\begin{eqnarray}
G_0' = ( f U_0 - G_{0(r)} t ) / \{ f t - f^2 U_0 ( G_{0(r)} / g_0^2 ) \}   ~.
\end{eqnarray}

Especially, when the composite velocity of relative motion is equal to zero, that is, $G_{0(r)} = 0$, the above will be reduced into, $G_0' = G_0$ , or
\begin{eqnarray}
V_0' + k_{rx} Y_0' = V_0 + k_{rx} Y_0   ~,
\end{eqnarray}
where $Y_0$ is the scalar quantity of the partial electromagnetic potential.

The above means that the speed of light, $V_0$ , is still variable, even if there is no composite velocity, $G_{0(r)}$ , of relative motion, between the two coordinate systems, $\mu$ and $\nu$. The speed of light, $V_0$ , is relevant to the scalar quantity of the partial electromagnetic potential. Next, when the composite velocity, $G_{0(r)}$ , of relative motion, is not equal to zero, the speed of light, $V_0$ , will deal with the scalar quantities of the partial electromagnetic and gravitational potential. In a general way, in case there are several relative motions in different directions, the speed of light, $V_0$, may associate with the partial potential of electromagnetic and gravitational fields, including their scalar quantities and vector quantities.

\subsection{Refractive index}

If we choose the quantity, $V_0$ , to be the speed of light in the vacuum, and $V_0'$ in the optical waveguide, it is capable of deriving the following refractive index, $n$, from the above,
\begin{eqnarray}
1 + k_{rx} Y_0' / V_0' = n + k_{rx} Y_0 / V_0' ~,
\end{eqnarray}
where $n = V_0 / V_0'$ .

Further, in case we select the quantity, $Y_0$ , to be zero in the vacuum, the above will be simplified into,
\begin{eqnarray}
n = 1 + k_{rx} Y_0' / V_0'  ~.
\nonumber
\end{eqnarray}

The above reveals that the refractive index, $n$, connects with not only the speed of light, but also the scalar quantity, $Y_0'$ , of partial electromagnetic potential. The refractive index is a function of the scalar quantity, $Y_0'$ , of partial electromagnetic potential, rather than the intrinsic property any more. In terms of various optical mediums, the remaining scalar quantities of partial electromagnetic potential within the mediums may be different from each other, so their refractive indices can be quite distinct accordingly. Obviously, the higher the partial electromagnetic potential $Y_0'$ is, the wider the rangeability of refractive index $n$ is. If the two coordinate systems have several relative motions in multi-directions, there may be more influencing factors (such as, velocity, field strength, and frequency and others) to impact the variation of refractive index.

\section{Experiment proposal}

In the flat composite space, $\mathbb{O}_U$ , described with the octonions, according to Eqs.(3) and (33), the octonion velocity is, $\mathbb{V} = \partial \mathbb{R} / \partial t$ , and the octonion partial potential is, $\mathbb{Y} = \partial \mathbb{X} / \partial t$ . The octonion composite velocity is written as, $\mathbb{G} = \mathbb{V} + k_{rx} \mathbb{Y}$. It means that the velocity $\mathbb{V}$ and partial potential $\mathbb{Y}$ both are able to exert some analogous influences on the space and time, in the octonion composite space.

Not only the Galilean-like transformation, Eqs.(61) and (62), but also the Lorentz-like transformation, Eqs.(63) and (64), reveal that the composite velocity, $g_{1(r)}$, of relative motion has an influence on the coordinate transformations between the two coordinate systems, in the flat composite space $\mathbb{O}_U$. This means that the velocity component and partial potential component in the composite velocity, $g_{1(r)}$ , will impact directly the Galilean-like transformation and Lorentz-like transformation. Apparently, it is able to modify and improve the Sagnac experiment and Aharonov-Bohm experiment, verifying these inferences with respect to the variable speed of light.

In the existing Sagnac experiments, the relative speed between the two coordinate systems will exert an influence on the coordinate transformation. According to the preceding analysis, the partial potential will also impact directly the coordinate transformation between the two coordinate systems. In other words, substituting the partial potential for the velocity, the movement of interference fringes must also be occurred in the Sagnac experiments. Consequently, the existence of partial electromagnetic potential can result in the movement of interference fringes, in terms of the Sagnac experiments without any relative speed in the rectilinear motion or curvilinear motion. Especially, we must contrast the Sagnac experiment subjected to the impact of partial electromagnetic potential with that free of the partial electromagnetic potential, distinguishing the movement of interference fringes. Up to now, the Sagnac experiments have never been verified under the influence of the partial electromagnetic potential, so the paper appeals intensely to validate a few relevant experiments.

In this experiment proposal, what we explore is that the influence of partial electromagnetic potential on the speed of light. Obviously, it is distinct from the influence of electromagnetic strength on the speed of light. As a result, the experiment proposal must isolate any interference of electromagnetic strength. We can learn the experimental methods relevant to the electromagnetic shielding from the Aharonov-Bohm experiments. In the latter, the electromagnetic strength is isolated effectively, while the contribution of the electromagnetic potential is emerged and measured adequately.

Referring to the experimental methods in the Aharonov-Bohm experiments, it is capable of inspecting the influence of the partial potential on the Sagnac experiments, analyzing the impact of the partial potential on the speed of light and refractive index.

\section{Conclusions and discussions}

In the octonion space $\mathbb{O}$ , from the properties of coordinate transformations, the scalar part of the octonion physical quantity is an invariant under some coordinate transformations, including the scalar part of octonion radius vector, scalar part of octonion velocity, and norm of octonion radius vector. A few combinations of these invariants will generate several basic postulates, such as Eqs.(8) and (9), for the Galilean transformation and Lorentz transformation in the octonion space. Further, other types of combinations may yield new basic postulates and coordinate transformations. For example, substituting the velocity for the radius vector in Eq.(6) may produce one new type of basic postulate and coordinate transformation.

Similarly, in the octonion composite space $\mathbb{O}_U$ , from the properties of coordinate transformations, the scalar part of the octonion composite physical quantity is an invariant under the coordinate transformations, including the scalar part of octonion composite radius vector, scalar part of octonion composite velocity, and norm of octonion composite radius vector. Some combinations of these invariants will produce several basic postulates, such as Eqs.(38) and (39), for the Galilean-like transformation and Lorentz-like transformation in the octonion composite space. Further, it is capable of inferring the coordinate transformation of the composite speed of light in the octonion composite space, explaining the reason for the variable speed of light (or refractive index) of the optical waveguides under the external fields.

By comparison with the Galilean-like transformation and Lorentz-like transformation, the octonion coordinate transformation, derived from the invariants in Eq.(39), is capable of explaining why the speed of light may be varied within the interface layer between two different optical waveguides under some circumstances, exploring some physical properties of refractive indices.

Making use of the properties of the octonion composite space $\mathbb{O}_U$ , it is able to deduce the relationships of spatial parameters and physical quantities in the curved spaces, generating a few inferences accordant with the GR (see Ref.[27]). In the vacuum with the external fields, it is capable of producing the correlations between the speed of light and the external field potential, achieving several inferences compatible with the refractive indices. In the octonion composite space $\mathbb{O}_U$, the inferences derived from Eq.(69) can be applied to analyze the physical phenomena with respect to the variations of the speed of light in the optical waveguides with the external fields, explaining why the refractive indices are variable. It should be noted that the partial electromagnetic potential, $( Y_0 / v_0 )$, possesses the dimension of the electromagnetic potential, but is not the electromagnetic scalar potential. So it can be considered as a part of the electromagnetic scalar potential, or one function of the electromagnetic potential. Obviously, various optical waveguides possess different values of the physical quantity $Y_0$ . Therefore, when the tiny beam of light transmits through some disparate optical waveguides, the speed of light must be transferred accordingly. It means that the variation of the electromagnetic scalar potential may cause the fluctuation of the speed of light in the composite space.

Because the composite velocity $G_{0(r)}$ of relative motion is not equal to zero, the variation of the speed of light in Eq.(68) will be more intricate than that in Eq.(69), observing several much more involved physical phenomena. Further, in case there are a few relative motions in multi-directions, it is necessary to take into account some situations, which are much more complicated than that in Eq.(65), measuring certain more convoluted physical phenomena. At that moment, the partial electromagnetic vector potential will also exert an impact on the variation of the speed of light. On the other hand, the partial gravitational potential makes a contribution to the composite velocity component $g_0$ , impacting the composite speed of light $G_0$ . In other words, the partial gravitational potential will impact directly the composite speed of light $G_0$, under certain circumstances. As a result, the composite speed of light, $G_0$ , will be variable within various gravitational fields, especially the ultra-strong gravitational fields in the astronomy. Even in the flat spaces, the emergence of the components of gravitational potential will still exert an influence on the composite speed of light, $G_0$, transforming the speed of light $V_0$ .

It should be noted that the paper explores merely some simple cases for the contributions of the radius vector and the integrating function of field potential on the coordinate transformations, in the octonion composite spaces. Despite its preliminary characteristics, this study can clearly indicate that the external electromagnetic potential and gravitational potential both may exert an impact on the speed of light in the optical waveguides. In the following study, it is going to further search for more influencing factors with respect to the multi-direction relative motions by means of the coordinate transformations theoretically. In the experiments, we shall survey the influences of the external electromagnetic potential and gravitational potential on the speed of light, probing deeply into several new properties of refractive indices in the optical waveguides.

\begin{acknowledgements}
This project was supported partially by the National Natural Science Foundation of China under grant number 60677039.
\end{acknowledgements}

\section*{Conflict of interest}
The author has no conflicts to disclose.

\section*{Data Availability}
The data that support the finding of this study are available within this article.

{}

\end{document}